\documentclass[12pt]{iopart}

\usepackage{graphicx}
\usepackage{xcolor}

\begin{document}

\title{Mesoscopic Möbius ladder lattices as non-Hermitian model systems}

\author{Jung-Wan Ryu}
\address{Center for Theoretical Physics of Complex Systems, Institute for Basic Science (IBS), Daejeon 34126, Republic of Korea}
\ead{jungwanryu@gmail.com}

\author{Martina Hentschel}
\address{Institute of Physics, Technische Universität Chemnitz, D-09107 Chemnitz, Germany}
\ead{martina.hentschel@physik.tu-chemnitz.de}


\begin{abstract}
While classic quantum chaos originated from the idea to set into context nonlinear physics and Hermitian quantum mechanics, non-Hermitian models have enhanced the field in recent years. At the same time, low-dimensional effective matrix models have proven to be a powerful tool in accessing the physical properties of a system in a semiquantitative manner. Here, we focus on two realizations of non-Hermitian physics in mesoscopic systems. First, we consider spiral optical microcavities in which the asymmetric scattering between whispering gallery modes induces the non-Hermitian behaviour. Second, for parity-time (PT) symmetric ladder lattices we compare circular and Möbius geometries. We find the effective coupling between even and odd parity modes to be symmetric but complex in a microscopically derived 2 $\times$ 2 matrix model, resulting in non-Hermitian behaviour as well. Most importantly, the Möbius topology acts like a scatterer that induces a qualitatively new form of (avoided) level crossing -- a PT-broken phase terminated by exceptional points --  resulting from the symmetric but non-Hermitian coupling. 
\end{abstract}

%
%
%
%
%

\section{Introduction}

The field of quantum chaos has deeply inspired more than a generation of physicists over the past three decades. Many of us, the student generation of the later 1990s, started off with the derivation of the Wigner surmise \cite{haake2001quchaos} of random matrix theory \cite{mehta_book_rmt}, impressed by how such a simple 2 $\times$ 2 matrix model, based on very general principles, could explain so many observations at once (that there was not a real proof for the generalisation to $N \times N$ matrices we were told, kept in mind, and most of us ignored for all practical purposes). 

The enormous multitude of examples in which 2$\times$2 matrices carry the essence of physics range from the Pauli spin matrices, via rotation and symmetry transformations in general, to effective models where the reduction of a complex system to an effective 2$\times$2 matrix helped to understand the kernel of a highly complex physical system by condensing it to four complex numbers constituting the four matrix entries.
Even beyond physics matrix models are widely used from game theory to chemistry \cite{hueckel1931}. 

In the cases we are interested in, the matrix $M$ has two diagonal elements $a$ and $d$, often the energies of the two level system, coupled by two off-diagonal elements $b$ and $c$. The richness of all the situations that can be described arises from the entries being real or complex numbers, and the symmetry properties of $M$ which can be (non-) Hermitian or (non-) symmetric. It directly reflects, and corresponds to, the underlying physical situation for which $M$ shall provide the effective Hamiltonian. 

A paradigm example, besides the Wigner surmise, is the avoided crossing of two energy levels, $E_1=a$ and $E_2=d$, that are coupled symmetrically by off-diagonal elements $b=\beta$ and $c=\beta^*$. The parameter $\beta$ quantifies the interaction between the two energy levels. The new energy levels $E_+, E_-$ in the presence of the perturbation $\beta$ that can, e.g., be induced by scattering, are well-known to read
\begin{equation}
    E_{\pm}= \frac{1}{2} (E_1 + E_2) \pm \frac{1}{2}\sqrt{(E_1-E_2)^2 + 4 |\beta|^2} \:.
    \label{eq_alc}
\end{equation}
The avoided level crossing is clearly seen at $\Delta E = E_1 - E_2 = 0$ where a level crossing accounts for zero interaction, or perturbation, $\beta$ = 0. As the perturbation is set on, a gap between $E_1$ and $E_2$ opens indicating the avoiding of the level crossing. 

Avoided level crossing will play a crucial role throughout this paper, and we shall see that a change from Hermitian symmetry to a non-Hermitian, parity-time (PT) symmetric \cite{Bender_2007} situation changes the level crossing scenario dramatically. Instead of the well-known and generic avoided level crossing feature, we will observe something qualitatively very different, namely that the two energy levels snap into a PT-symmetry broken phase with the same real parts of the energy and opposite imaginary parts, confined by an exceptional point on either side. 

In the present paper  we will discuss the paradigmatic example of avoided level crossing mainly into two directions. 
In the following Section \ref{sec_effmodels} we will highlight results from 2 $\times$ 2 matrix models that proved to be particularly useful for mesoscopic physics and quantum chaos \cite{stoeckmann_1999}, thereby focussing on asymmetric coupling in optical microcavities \cite{Vahala_book}. In Section \ref{sec_ladderlattices} we introduce the PT-symmetric circular and Möbius-type ladder lattice models and discuss and illustrate the peculiar manner in which topology affects level crossing scenarios as it plays the role of an effective perturbation to the system. \textcolor{black}{Whereas this model was introduced in Ref.~\cite{PTladder_PRA}, we provide here a detailed qualitative explanation in terms of effective 2$\times$2 and 4$\times$4 matrix models that allow for a deeper physical understanding. We add the Hermitian case into the discussion and compare it with the non-Hermitian situation and with effective models used in other fields such as chemistry.} \textcolor{black}{Moreover,} we discuss the role of bulk vs.~boundary effects and correspondences between Möbius-type ladder lattice models and chaotic billiards in Section \ref{sec_discussion} and close with a brief summary and outlook towards mesoscopic optics. 

\section{Achievements of effective 2 $\times$ 2 matrix models: Chirality in optical microcavities}
\label{sec_effmodels}

\begin{figure}
    \centering
    \includegraphics[width=6cm]{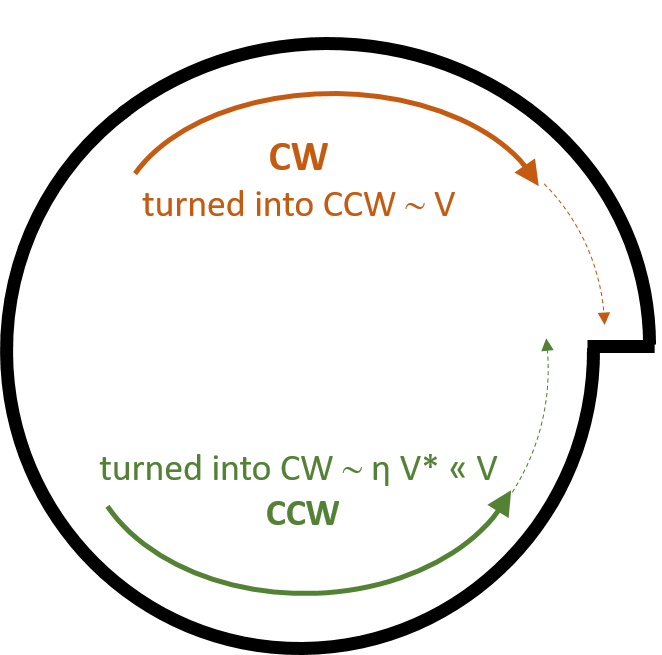}
    \caption{Asymmetric scattering between clockwise (CW) and counterclockwise (CCW) travelling waves in an optical microcavity induced by the asymmetry of resonator geometry, here illustrated by the case of a spiral. The resonance dynamics can be effectively described by a suitable 2 $\times$ 2 matrix such as Eq.~(\ref{H_1}) that confirms the dominance of the CCW wave as it is less affected by scattering into the oppositely travelling mode.}
    \label{fig:2by2_optics}
\end{figure}

To illustrate the success and achievements of simple matrix models, we pick an example from the field of quantum chaos in optical microcavities which proved to be a versatile and rich mesoscopic model system 
\cite{RMP_cao_wiersig,QuaChaCav:Nockel:1997,QuaChaCav:Gmachl1998,PhysRevLett.127.203901,yfxiao2012,vollmer_xiao,2d_microcavitylasers}.
To this end we consider asymmetric coupling that can be straightforwardly realized in deformed optical microcavities with a lack of mirror symmetry. A well-known example is a spiral microcavity \cite{nonorth_spiral,nonorth_generic} shown in Fig.~\ref{fig:2by2_optics}. It was shown that this induces a chirality into the eigenmodes that, at the same time, become nonorthogonal. 



The effective, toy-model like Hamiltonian can be written as 
\begin{equation}
H_{\mathrm{ch}} = \left( 
\begin{array}{c c}
E_0 & 0 \\
0 & E_0 
\end{array}
\right)
+ 
\left( 
\begin{array}{cc}
\Gamma & V \\
\eta V^* & \Gamma 
\end{array}
\right) \:.
\label{H_1}
\end{equation}
The first matrix describes uncoupled clockwise (CW) and counterclockwise (CCW) modes with the same energy $E_0$, such as whispering gallery modes in a disk for which the second matrix would not exist. So the second term can be interpreted as describing the scattering between CW and CCW modes that occurs in deformed disk microcavities. While $\Gamma$ are the total scattering rates and assumed to be equal for simplicity, the off-diagonal elements describe the scattering between CW and CCW rotating modes and can be different. While the complex number $V = |V| \, e^{i \theta}$ describes the scattering from CW to CCW traveling waves, $\eta \, V^*$ describes the opposite process from CCW to CW travelling waves. Here, we assume a geometry such that this process is  weaker, $ 0 \leq\eta <1$, as illustrated in Fig.~\ref{fig:2by2_optics}. For example, it would be the case in a spiral cavity with a notch that hinders the propagation of CW traveling modes \cite{nonorth_spiral}.

This eigenvalue problem can be easily solved and yields eigenvalues
\begin{equation}
E_\pm = E_0 + \Gamma \pm \sqrt{\eta} |V| 
\label{eq_En_spiral}
\end{equation}

and right eigenvectors
\begin{equation}
{\mathbf a_\pm} = \frac{1}{\sqrt{2}} 
\left( \begin{array}{c} 
1 \\ \pm \sqrt{\eta} \, e^{-i \theta}
\end{array} \right)
\end{equation}

The structure of the eigenvector directly reveals the dominance of the upper, namely the CCW component (weight $\sim 1$) over the CW traveling component with a weight $\sim \eta < (\ll) 1$. This effective Hamiltonian well describes the chiral modes in fully asymmetric microcavities such as spiral shaped microcavities and circular microcavities with scatterers \cite{nonorth_spiral,nonorth_generic,PhysRevA.84.063828}.

We note that the non-reciprocal (asymmetric) toy-model like Hamiltonian, Eq.~(\ref{H_1}), can be transformed into a reciprocal (symmetric) Hamiltonian by a proper similarity transformation \cite{Miri_2016}. In spite of the mathematical analogy between the two Hamiltonians, they describe very different physical systems because of different basis modes. For example, while Eq.~(\ref{H_1}) describes chiral modes separated by oppositely signed angular momenta, the transformed one describes modes localized in distinct regions in real space \cite{ep_pra2009}. 

In the following, we will leave quantum chaos in photonic systems behind and switch to another non-Hermitian system, namely the PT-symmetric Möbius strip. To this end we will induce non-Hermiticity in a qualitatively different way by considering complex eigenenergies (i.e.~complex diagonal entries in the $2 \times 2$ matrix model) such that parity-time (PT) symmetry is respected. At the same time, we will consider symmetric coupling and thus standing wave modes (equal contributions from $\pm m$ angular momenta) instead of travelling wave modes (one $m$-sign dominates) as discussed above. We will leave for now the optical microcavity systems but not without having pointed out the richness of exceptional points that can be realized in these systems, see e.g.~\cite{kullig_EP3,WiersigYang_Science,Wiersig:20,PhysRevA.78.015805,PhysRevA.96.063823,2016Sci...353..464M,Bosch2019,MiriAlu,541852d425a547d591a9f10f3f1804c3} and references therein. 

\section{PT-symmetric mesoscopic ladder lattices}
\label{sec_ladderlattices}

\begin{figure}[tb]
 	\centering
\includegraphics[width=14cm]{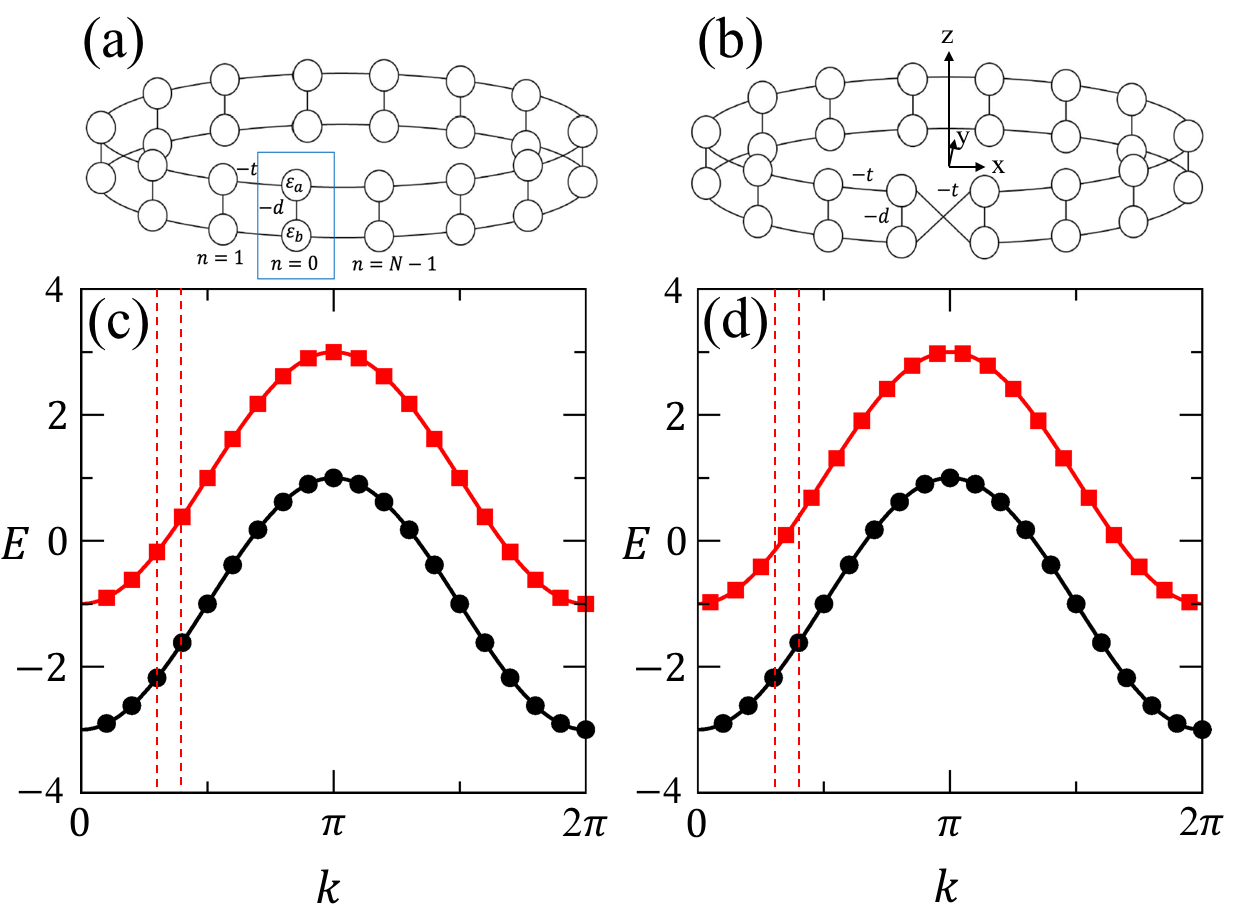}
\caption{(a) A circular ladder lattice (CLL) and (b) a M{\"o}bius ladder lattice (MLL) with $N=20$ sites. The blue box represents the unit cell of the CLL, which has two sites with intra-unit cell hopping strength $d$ and inter-unit cell hopping strength $t$. 
(c) Bands of 
eigenenergies of the CLL with 20 unit cells. (d) Bands 
and eigenenergies of the MLL with 20 unit cells. Red and black lines represent the bands of states with odd and even parities, respectively, and red squares and black circles represent the eigenenergies of states with odd and even parities, respectively. The dotted vertical lines illustrate different $k$ values and point out the difference between the CLL and MLL.
}
\label{fig_band}
\end{figure}

\begin{figure}
    \centering
    \includegraphics[width=16 cm]{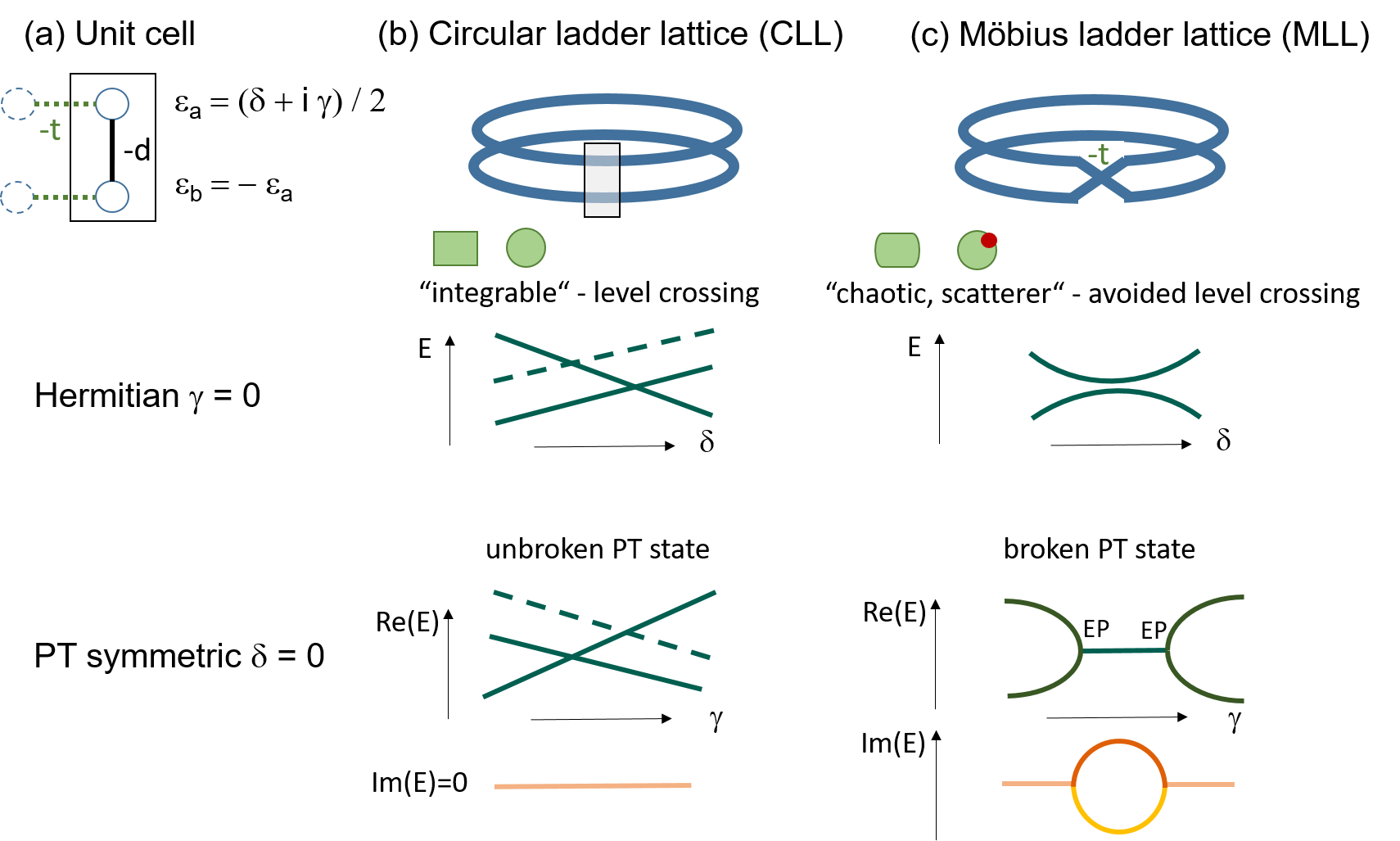}
    \caption{Schematics of the relation between symmetries and level crossing behaviour for Hermitian and PT-symmetric ladder lattices. (a) Unit cell of the ladder lattice with onsite energies $\varepsilon_{a,b}$. (b) CLL in the Hermitian ($\gamma=0$) and PT-symmetric ($\delta=0$) situation. Both cases show level crossings reflecting the high, unbroken symmetry of the system that is also found in regular billiards such as the rectangle or disk (inset). The slope of the branches and therefore the intersection points differ in both cases according to Eq.~(\ref{eq_band}). (c) Same for an MLL where the twist acts as a scatterer, implying a chaotic system similar to the situation in billiards (inset). This yields avoided level crossing in the Hermitian case. However, in the non-Hermitian case a PT-broken state (with non-vanishing imaginary energy parts) forms in between two exceptional points (EP), representing a new form of level crossing behaviour.}
    \label{fig_scheme}
\end{figure}

The effective $2\times2$ matrix model of the previous section, Eq.~(\ref{H_1}), can be extended to capture other system classes. Here, we are interested in non-Hermitian one dimensional (1D) lattice models with a 2-sites unit cell as shown in Figs.~\ref{fig_band} and \ref{fig_scheme}(a) . The motivation for this is the twist that was originally inspired by optical Möbius cavities \cite{moebiusEPL} where the light collects geometric phases. In a ladder lattice, the Möbius twist acts rather differently, as we will discuss now.

The on-site potentials of the two sites in a unit cell are $\varepsilon_a$ and $\varepsilon_b$, and the intra- and inter-unit cell hopping strengths are $d$ and $t$, respectively. The effective Hamiltonian of an infinite 1D ladder lattice can be expressed as \cite{PTladder_PRA}
\begin{equation}
H_{\mathrm{ll}} = \left( 
\begin{array}{c c}
\varepsilon_a - \textcolor{black}{2} t \cos{k} & -d \\
-d & \varepsilon_b - \textcolor{black}{2} t \cos{k} 
\end{array}
\right)\:,
\end{equation}
where $k$ is the Bloch wave vector. The on-site potentials can be chosen to be antisymmetric without loss of generality, i.e., $\varepsilon_a = -\varepsilon_b = \delta/2+i\gamma/2$. We can consider the Hermitian ($\gamma = 0$) and PT-symmetric ($\delta = 0$) cases as limiting situations. 
The eigenvalues and eigenstates of this Hamiltonian are
\begin{eqnarray}
\label{eq_band}
E_{\pm} = - 2 t \cos{k} \pm \sqrt{d^2 + \left(\frac{\delta}{2}+i\frac{\gamma}{2}\right)^2}, \\
\psi_{\pm} =
\left( 
\begin{array}{c}
-\left(\frac{\delta + i \gamma}{2d}\right) \mp \sqrt{1+ {\left(\frac{\delta + i \gamma}{2d}\right)}^2}  \\
1 
\end{array}
\right),
\label{eq_es}
\end{eqnarray}
where $E_{\pm} (k)$ represent two parallel energy bands 
as shown in Fig.~\ref{fig_band}(c,d). 
In a ladder lattice with symmetric on-site potential, i.e., $\varepsilon_a = -  \varepsilon_b = (\delta + i \gamma)/2 = 0$, the energy bands 
form two sub-bands corresponding to states with odd 
parity (upper band, $E_+$) and even
parity (lower band, $E_-$), cf.~Fig.~\ref{fig_band}.
As $d$ and $t$ increase, the distance between the two bands and the band widths increase, respectively.

We start by considering a CLL with $N$ unit cells under periodic boundary conditions. These imply that the Bloch wave vector $k$ is the same as that in a single ring lattice, since 
\begin{equation}
\left( 
\begin{array}{c}
u_{a} (0) \\
u_{b} (0) 
\end{array}
\right)
=
e^{i k N}
\left( 
\begin{array}{c}
u_{a} (N) \\
u_{b} (N) 
\end{array}
\right)\:.
\label{bc_CLL}
\end{equation}
Here $u(x)$ is a periodic function of position $x$, which satisfies $u(x+ \rho)=u(x)$ when the potential is a periodic function with period $\rho$, i.e., we write the Bloch wavefucntion of a periodic system as $\psi(x) = e^{i k x} u(x)$. Assuming periodic boundary conditions, the eigenvalues of the Hamiltonian of a CLL with $N$ unit cells are 
\begin{equation}
    E_{\pm} = - 2 t \cos{\frac{2 n \pi}{N}} \pm d ~~~ (n=1,\cdots,N) \:,  
\end{equation}
where $E_{\pm}$ are the eigenvalues of the corresponding eigenstates with odd and even parities [Fig.~\ref{fig_band} (c)]. They possess an equidistant spacing of $\Delta k = 2 \pi /N$, like those in single ring lattices. The eigenvalues locate on the energy bands $\varepsilon_{\pm} = - 2 t \cos{k} \pm d$ with $0 \leq k \leq 2 \pi$.

Next, the eigenenergies in the MLL can be obtained by replacing the periodic boundary condition for a CLL from Eq.~(\ref{bc_CLL}) with the boundary condition for an MLL as follows:
\begin{equation}
\left( 
\begin{array}{c}
u_{a} (0) \\
u_{b} (0) 
\end{array}
\right)
=
e^{i k N}
\left( 
\begin{array}{c}
u_{b} (N) \\
u_{a} (N) 
\end{array}
\right)\:.
\label{bc_MLL}
\end{equation}
Under this twisting boundary condition, the eigenvalues of the Hamiltonian of an MLL are 
\begin{eqnarray}
    E_{-} &= -2 t \cos{\frac{2 n \pi}{N}} - d ~~~ & \mathrm{for~even~parity} \:,\\
    E_{+} &= -2 t \cos{\frac{(2 n - 1) \pi}{N}} + d ~~~ & \mathrm{for~odd~parity} \:,
\end{eqnarray}
where $n = 1, . . . , N$. The Bloch wave vectors and corresponding eigenenergies are the same as those in the CLL in the case of even-parity eigenstates, but different in the case of odd-parity eigenstates [Fig.~\ref{fig_band} (d)]. The upper and lower sites of the unit cell exhibit opposite signs of amplitudes in the case of odd-parity eigenstates (thus allowing to define an orientation), 
while the sites have the same signs of amplitudes (meaning no orientation) in the case of even-parity eigenstates. Consequently, for the even-parity case, there is no effect of breaking orientation in the MLL since there is no orientation present already in the CLL. This changes when odd-parity states are considered for which deviations from the CLL are expected and observed, cf.~Fig.~\ref{fig_band}(d). It is noted that the symmetry of the unit cell parameters plays an important role in obtaining the eigenenergies.

 With respect to the symmetries present, it is useful to explore the relation of the CLL and MLL to two dimensional (2D) circular cavities without and with a defect or a scatterer. First, we consider symmetries of the CLL in comparison with a 2D 
 disk cavity in the ($x$,$y$) plane (see Fig.~\ref{fig_band}(b) for the definition of the coordinate system). 
 Both systems 
 possess (discrete) rotational symmetry in the ($x$,$y$) plane. The CLL has additional mirror symmetry about the $z$-axis when $\varepsilon_a = \varepsilon_b = 0$ that is not relevant for the 2D disk. Consequently, the energy spectra in a CLL show a degeneracy of energies of $k$ and $2\pi - k $ corresponding to the degeneracy with respect to angular momentum $\pm m$ between CW and CCW travelling WGMs in a circular cavity. 
 Additionally, the CLL spectra distinguish even $(u_a = u_b)$ and odd $(u_a = - u_b)$ parity states with respect to the mirror symmetry about $z$-axis of the unit cell. These degeneracies and parities can be confirmed in Eqs.~(\ref{eq_band}) and (\ref{eq_es}). Even if $\delta$ or $\gamma$ increases, the 
 degeneracy of the CW and CCW states is preserved by unitary transformations with rotation angles being a function of $\delta$ or $\gamma$, though the mirror symmetry about $z$-axis is broken \cite{PTladder_PRA}.

In turn, an MLL with a twisted part can be considered as a circular cavity with a defect or a scatterer as a perturbation (both placed at the $y$ axis). The rotational symmetries of the CLL and a circular cavity are broken by the twisted part and the defect, respectively. Consequently, both systems have now discrete mirror symmetries about the $y$-axis. The (originally degenerate) $\pm k$ states, i.e. the CW and CCW rotating modes, will split into states with even and odd parities about the $y$-axis which defines the axial symmetry, cf.~the dashed lines in Fig.~\ref{fig_band}(d). 
Such split states resulting from axial mirror symmetries are common in deformed optical microcavities with symmetry axes 
\cite{stoeckmann_1999,PhysRevA.70.023809,PhysRevA.72.023815,PhysRevLett.91.073903,PhysRevLett.100.033901,annbill}.

It is of particular interest to study the level crossing dynamics 
in the presence of the different symmetries when the symmetry of the unit cell ($\varepsilon_{a}= \varepsilon_b=0$) is broken, that is, when $\delta$ and $\gamma$ differ from zero. 
To this end we start from Fig.~\ref{fig_band} ($\delta= \gamma=0$) and pick $k$ values that yield energies $E$ around $E=0$. Before we discuss the exact results in Fig.~\ref{fig_spectra} below, let us consider possible scenarios in a qualitative manner as sketched in Fig.~\ref{fig_scheme}. 

In a CLL, 
Fig.~\ref{fig_scheme}(b), lifting the degeneracy of the onsite energies $\varepsilon_{a,b}$ by detuning either $\delta$ or $\gamma$ from zero yields level crossings. The only difference between the Hermitian and PT-symmetric cases is the slope of the energy levels that is opposite when $\delta$ or $\gamma$, respectively, are changed, in agreement with Eq.~(\ref{eq_band}). The situation changes in an MLL where the twist can be considered as a scatterer, i.e.~as an element inducing an interaction into the system, thus a coupling between the energy levels. In the Hermitian case, this is the prototype of avoided level crossing and captured by Eq.~(\ref{eq_En_spiral}) above by lifting the degenerate energy value $E_0$ to become $E_1$ and $E_2$  (setting $\Gamma=0$ and $\eta=1$ while $\delta$ or $\gamma$ takes the role of $(E_1 - E_2)/2$).
The new eigenenergies follow the well-known formula Eq.~(\ref{eq_alc}), $(E_1 + E_2) /2 \pm \sqrt{(E_1 - E_2)^2 /4 + V^2}$ showing the avoided crossing as function of $(E_1 - E_2)/2$.

The presence of PT symmetry allows for complex energies, and that adds a very different and qualitatively new scenario in the regime $\gamma < 2$, i.e.~below the phase transition point where PT symmetry 
is $\textit{not}$ broken. 
It is illustrated in the lower right panel of Fig.~\ref{fig_scheme}. The avoided level crossing is replaced by a range of parameters $\gamma$ where the PT symmetry actually \textit{is} broken: the eigenenergies are complex with degenerate real parts and opposite imaginary parts. This PT-broken region is terminated by an exceptional point (EP) on either side \cite{PTladder_PRA}.
We can model this behaviour by an (effective) 2$\times$2 matrix as in the Hermitian case by adding a new parameter $\xi$ that modifies the coupling between the two states such that it represents the Möbius topology. 
The coupling can be deduced from the $N \times N$ tight-binding model of the ladder lattice with $N$ sites. The amplitude equations of the $n$-th unit cell with neighboring unit cells $n-1$ and $n+1$ can be taken out and read
\begin{eqnarray}
\label{CLL_Eqs1}
    E a_n = \varepsilon_a a_n - d b_n - t a_{n-1} - t a_{n+1} \\
    E b_n = \varepsilon_b b_n - d a_n - t b_{n-1} - t b_{n+1},
\end{eqnarray}
where $a_n$ and $b_n$ represent upper and lower sites of the $n$-th unit cell, respectively, and $E$ is the eigenenergy. This can be rewritten 
in terms of a basis of even and odd parity states with amplitudes $f_n$ and $p_n$, respectively as
\begin{eqnarray}
\label{CLL_Eqs2}
    E f_n = (\varepsilon_{+} + d) f_n + \varepsilon_{-} p_n - t (f_{n-1} + f_{n+1}) \\
    E p_n = (\varepsilon_{+} - d) p_n + \varepsilon_{-} f_n - t (p_{n-1} + p_{n+1}),
\end{eqnarray}
where $\varepsilon_{\pm}=(\varepsilon_a \pm \varepsilon_b) / 2$, $f_n = (a_n - b_n)/\sqrt{2}$, and $p_n = (a_n + b_n)/\sqrt{2}$.

So far, we have stuck to the ordinary periodic boundary conditions realized in the CLL. If we now include the twisted boundary conditions between the ($n-1$)th and $n$-th unit cell that define the topology of the MLL, one of the $t$ couplings has to change its sign (directly reflecting the exchange of $a_{n-1} \leftrightarrow b_{n-1}$ at the twist), so the equations are changed into
\begin{eqnarray}
\label{CLL_Eqs3}
    E f_n = (\varepsilon_{+} + d) f_n + \varepsilon_{-} p_n + t f_{n-1} - t f_{n+1} \\
    E p_n = (\varepsilon_{+} - d) p_n + \varepsilon_{-} f_n - t p_{n-1} - t p_{n+1},
\end{eqnarray}
We can directly read-off the coupling represented by the off-diagonal terms to be equal to $\varepsilon_{-}= \delta + i \gamma$, so we find a symmetric but non-Hermitian coupling \cite{gemming_private}. Note that its origin are the complex on-site energies of the unit cell that take the character of (imaginary) hoppings when we rotate the basis.



The resulting effective 2$\times$2 matrix and eigenenergies $\lambda_{1,2}$ for such a symmetric but non-Hermitian coupling can be written as
\begin{eqnarray}
    H_{\mathrm{eff}} &=& \left( \begin{array}{cc}
        E_1 & \gamma + i \xi \\
        \gamma + i \xi & E_2
          \end{array} \right)
     \:, \\  
     \lambda_{1,2} &=& \frac{E_1 + E_2}{2} \pm \sqrt{\frac{(E_1 - E_2)^2}{4} + (\gamma + i \xi)^2} \:.
        \label{eq_alcpt}
\end{eqnarray}
The additional term $ i \xi$ under the square root ensures the desired behaviour, namely allowing the \textcolor{black}{radicand} of the square root to become \textcolor{black}{complex}. 
We note that an asymmetric coupling ($\gamma \pm \xi$) as we know it from the spiral cavity, \textcolor{black}{cf.~Eq.~\ref{H_1}}, would fulfill the same purpose, however, is not realized here. 


\begin{figure}[tb]
 	\centering
\includegraphics[width=14cm]{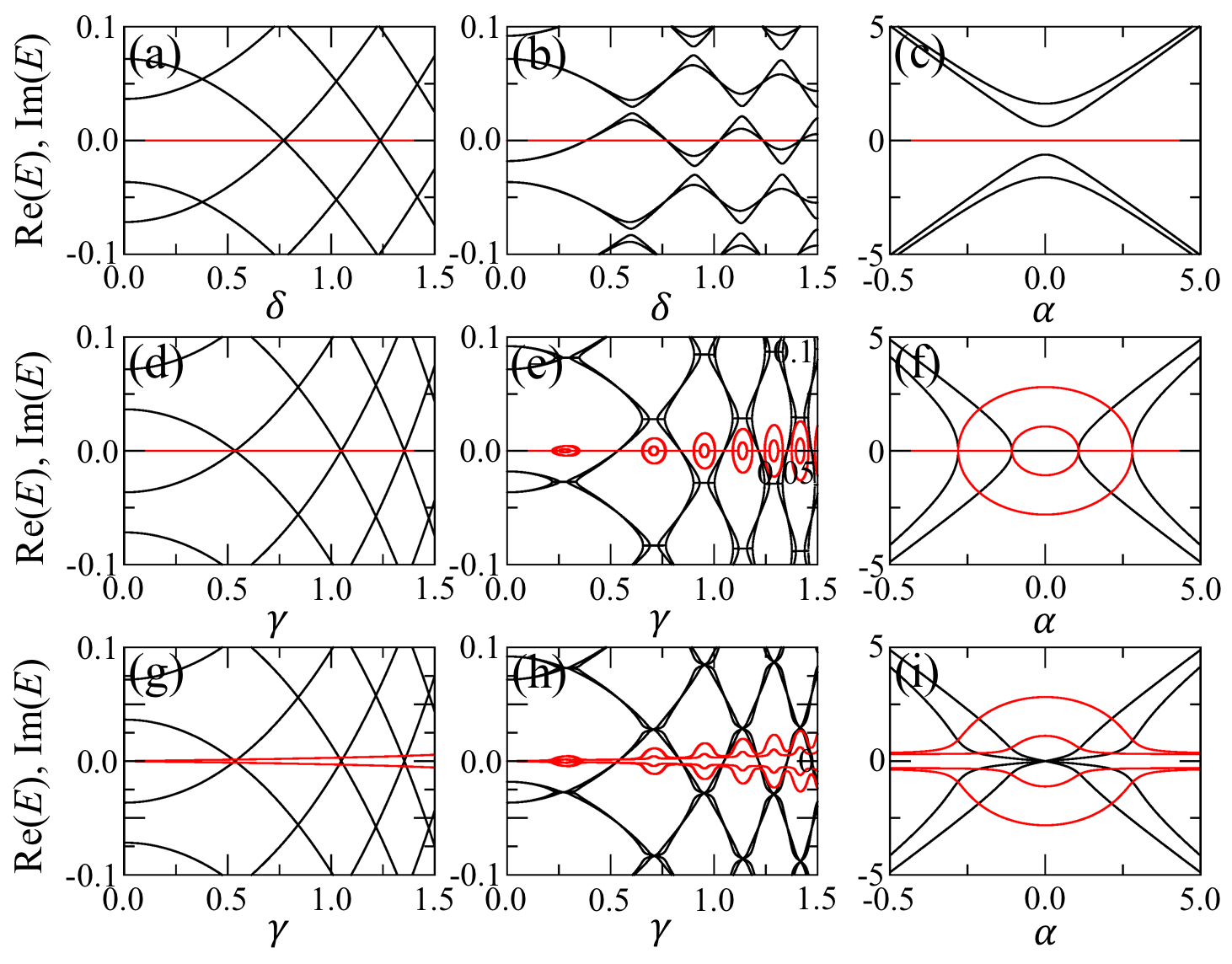}
\caption{
Real (black) and imaginary (red) eigenenergies as a function of $\delta$ in Hermitian (a) CLL and (b) MLL with $100$ unit cells when $d=t=1$ and $\gamma=0$. Real (black) and imaginary (red) eigenenergies as a function of $\gamma$ in PT-symmetric (d) CLL and (e) MLL with $100$ unit cells when $d=t=1$ and $\delta=0$. Real (black) and imaginary (red) eigenenergies as a function of $\gamma$ in non-Hermitian (non-PT-symmetric) (g) CLL and (h) MLL with $100$ unit cells when $d=t=1$ and $\delta=0.5$. Real (black) and imaginary (red) eigenenergies for the 4 $\times$ 4 matrix model in Eq.~(\ref{4by4mat}) as a function of $\alpha$ when (c) $\xi = 0.0$, (f) $\xi = 2.0$, and (i) the generic case with $\xi = 2.0$ and $\alpha \rightarrow \alpha + 0.3 i$ with $\beta = 1.0$. 
}
\label{fig_spectra}
\end{figure}

One feature that we have not yet considered is the degeneracy of the energy levels involved which is crucial in the real ladder models. This is taken into account in
Fig.~\ref{fig_spectra} which shows the resulting energy spectra for the CLL and MLL in the Hermitian and PT-symmetric cases. 
In the Hermitian CLL, Fig.~\ref{fig_spectra}(a), as $\delta$ increases, the energy of the lower energy band (cf.~Fig.~\ref{fig_band}(c)) with even parity Bloch states decreases, while the energy of the upper energy band with odd parity states increases because the band gap increases. 
The level crossings are evident. 
In contrast to the Hermitian case, in the PT-symmetric CLL, Fig.~\ref{fig_spectra}(d), the energy of the lower energy band (cf.~Fig.~\ref{fig_band}(c)) with even parity Bloch states increases  as $\gamma$ increases, while the energy of the upper energy band with odd parity states decreases because the band gap of real parts of energies decreases 
in the region of PT-symmetric phases where $E_{\pm}$ takes real values. This results in level crossings as well, however with a different pattern  (cf.~Fig.~\ref{fig_spectra}(a)). 


While the level degeneracy does not affect the CLL spectra, it becomes crucial in the MLL as we shall see now, Fig.~\ref{fig_spectra}(b, e). 
In the Hermitian MLL, Fig.~\ref{fig_spectra}(b), energy level crossings of four states in the CLL change into pairs of avoided level crossings with slightly different minimal energy spacing that occur at the same value of $\delta$. 
Analogously, two pairs of energy levels are visible in the PT-symmetric MLL, 
Fig.~\ref{fig_spectra}(e) which form now, however, a PT-broken state terminated by two EPs. These changes are caused by the non-orientability of the MLL resulting from the Möbius twist that breaks the rotational symmetry. 
In contrast to this, the CLL keeps its orientability in both the Hermitian and the PT-symmetric case. We note that the energy gaps of avoided crossings in Hermitian MLLs increase as $\delta$ increases,  while the PT-broken state range 
between pairs of EPs 
decreases as $\gamma$ increases in the PT-symmetric MLLs. 
In the generic situation of non-Hermitian and non-PT-symmetric ladder lattices, there is no PT-phase transitions as a function of $\gamma$, see Fig.~\ref{fig_spectra} (g, h) where $\delta=0.5$.


An explanation of the observed behaviour in terms of an effective matrix model which has now to be based on a $4 \times 4 $ matrix, is straightforward in the Hermitian situation. It is actually well-known in a very different field that we are happy to mention here in the context of effective low-dimensional matrix models, namely the Hückel model for molecular orbital theory in chemistry \cite{hueckel1931}. It was introduced in 1931 by Erich Hückel in order to explain the energy levels of the benzene ring. Here, we apply it to a slightly different molecule - after all, we do not want a 6 $\times$ 6, but rather a 4 $\times$ 4 matrix. The molecule that serves our purposes is butadiene, CH$_2$ = CH - CH = CH$_2$. Using the Hückel approximation for the Hamilton operator in the basis of the $p_z$ orbitals, the corresponding  (so-called inter-molecular connection or topological) matrix can be used to obtain its eigenenergies $E_i$ from the resulting secular equation with on-site energies $\alpha$ and couplings $\beta$; we would like to refer the interested reader to \cite{2020Butadiene} for details:
\begin{equation}
 \left|
 \begin{array}{cccc}
  \alpha - E  & \beta & 0 & 0\\
  \beta &  \alpha - E  & \beta & 0 \\
   0 &  \beta &  \alpha - E  & \beta  \\
   0 & 0 & \beta &  \alpha - E 
\end{array}
\right| = 0 \:.
\label{eq_butadien}
\end{equation}
Introducing the variable $x = ( \alpha - E ) / \beta$ simplifies things further and yields solutions 
\begin{equation}
    x^2 = \frac{3 \pm \sqrt{5}}{2} \nonumber
\end{equation}
corresponding to (approximately) $ x = \pm 0.618$ and $ x= \pm 1.618$. The four eigenvalues are symmetrically arranged about $\alpha$ and possess two different spacings, namely
\begin{eqnarray}
    E_1 & = & \alpha + 1.618 \beta  \nonumber \\
    E_2 & = & \alpha + 0.618 \beta \nonumber \\
    E_3 & = & \alpha - 0.618 \beta \nonumber \\
    E_4 & = & \alpha - 1.618 \beta \nonumber \:.
\end{eqnarray}
This is exactly the behaviour that we see in Fig.~\ref{fig_spectra}(b) at the (avoided) degeneracy points. Extension to their vicinity is straightforward along the lines of the 2$\times$2 matrix model for avoided level crossings, with applied stress being one of the relevant external parameters in the chemical context.
We mention that the chemical orbitals are obtained as superposition of the resulting eigenstates classified with respect to the number of nodes \cite{2020Butadiene}.


Based on this insight and the discussion above, Fig.~\ref{fig_spectra}(e) is readily understood as the generalisation of two avoided level crossings with different minimal energy spacings to the non-Hermitian, PT-symmetric situation. Similar to the two avoided level crossings in the Hermitian case, we find a pair of PT-broken states each sandwiched between two EPs. 
This behaviour is readily cast into an effective 4 $\times$ 4 matrix model describing two unit cells (in parity basis) and their mutual coupling.
Here, we use the asymmetric coupling that captures the asymmetry induced in the hopping of even and odd states, respectively, across the Möbius twist scatterer. For simplicity, we assume the same asymmetry for all off-diagonal matrix elements as we had set $d=t=1$, 


\begin{equation}
\label{4by4mat}
H_{\mathrm{4\times4}} = \left( 
\begin{array}{cccc}
  \alpha  & \beta + \xi & 0 & 0\\
  \beta - \xi &  - \alpha  & \beta + \xi & 0 \\
   0 & \beta - \xi &  \alpha  & \beta + \xi \\
   0 & 0 & \beta - \xi &  - \alpha 
\end{array}
\right) \:\:. \:\:
%
\end{equation}
It is Hermitian when $\xi=0$ and non-Hermitian when $\xi \neq 0$, thus effectively describing PT-symmetric models. A pair of avoided crossings and two pairs of PT-phase transitions appear when $\beta > \xi$ and $\beta < \xi$, respectively, as $\alpha$ increases (see Fig.~\ref{fig_spectra}(c,f)). When $\beta < \xi$, $\alpha$ 
can be considered as an effective imaginary on-site potential in MLLs. If we add the imaginary parts to $\alpha$, i.e., $\alpha \rightarrow \alpha + 0.3 i$, the generic non-Hermitian and non-PT-symmetric situation (Fig.~\ref{fig_spectra}(h)) is realized and reproduced, as shown in Fig.~\ref{fig_spectra}(i).


\section{Discussions}
\label{sec_discussion}

\subsection{Boundary vs.~bulk effects}
\label{sec_bulkboundary}

\begin{figure}[tb]
 	\centering
\includegraphics[width=14cm]{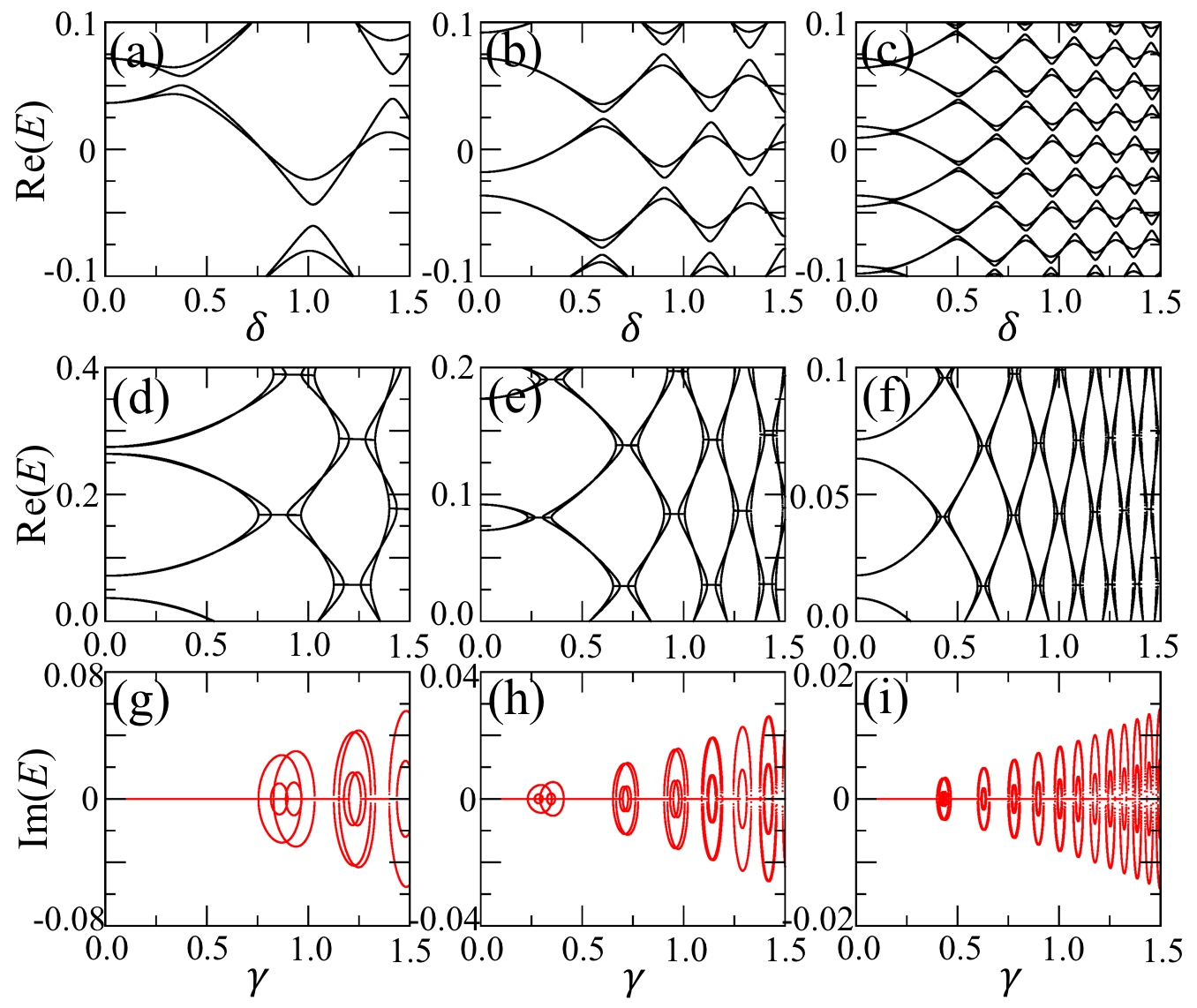}
\caption{Selected real eigenenergies as a function of $\delta$ in a Hermitian MLL with (a) 50, (b) 100, and (c) 200 unit cells when d = t = 1. Real (black) and imaginary (red) parts of the selected eigenenergies as a function of $\gamma$ in a PT-symmetric MLL with (d, g) 50, (e, h) 100, and (f, i) 200 unit cells when d = t = 1.
}
\label{fig_size}
\end{figure}

\begin{figure}
    \centering
    \includegraphics[width=12cm]{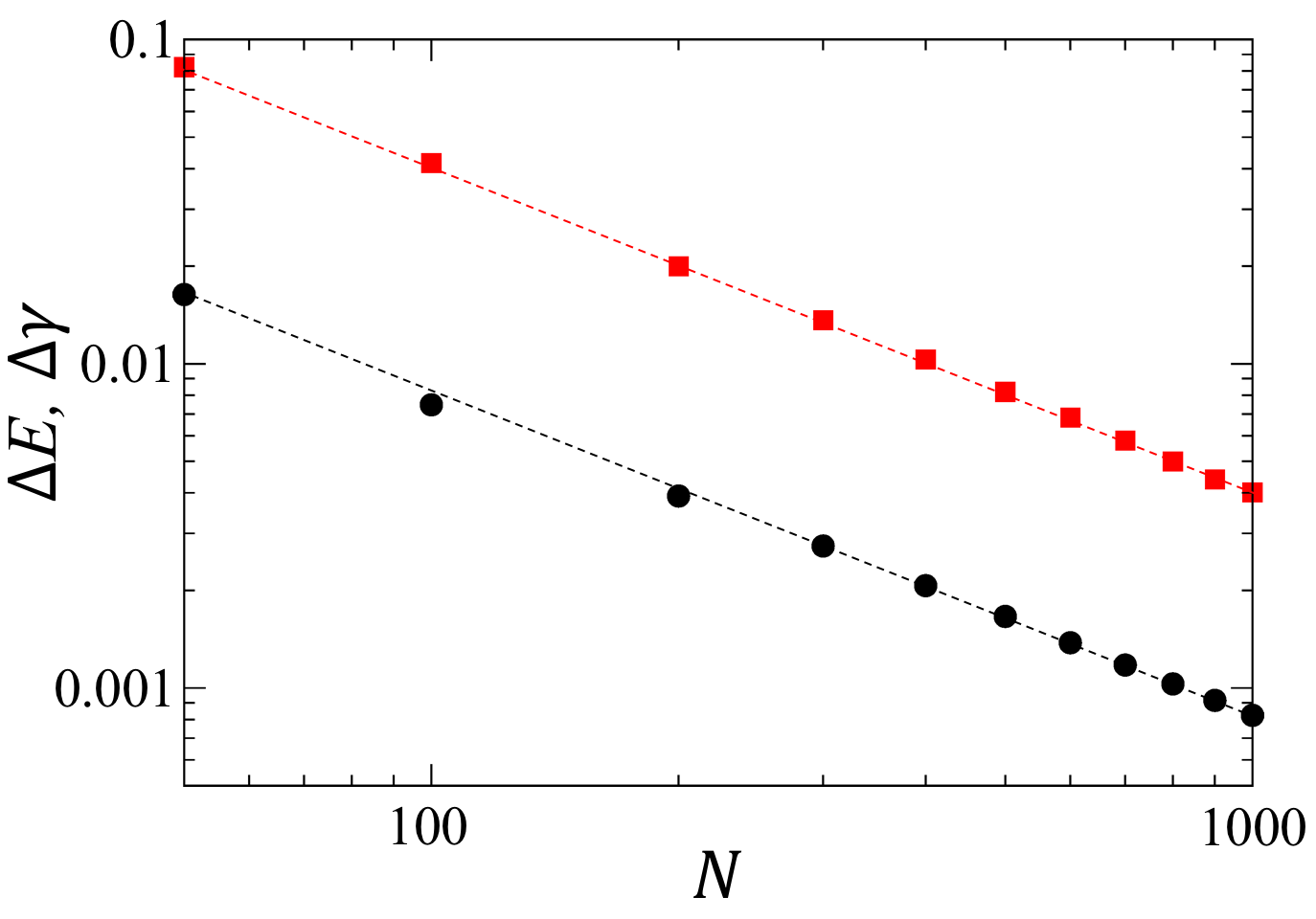}
    \caption{Energy gap $\Delta E$ (black circles) in Hermitian MLL and parameter gap $\Delta \gamma$ (red rectangles) in PT-symmetric MLL as a function of system size $N$. The dashed lines denote that $\Delta E$ and $\Delta \gamma$ are inversely proportional to $N$, i.e., $\propto 1/N$.}
    \label{fig_gapsize}
\end{figure}

We now return to our ladder lattice systems and discuss the influence of the system size on our findings. The results are shown in Fig.~\ref{fig_size} and Fig.~\ref{fig_gapsize}. 
Figure~\ref{fig_size} shows the increase in the number of states and the decrease in the energy and parameter gap sizes as the bulk size $N$ of unit cells increases while the boundary conditions are not changed. 
Figure~\ref{fig_gapsize} shows that energy and parameter gaps which are properly selected around zero energy are inversely proportional to the system size, $\Delta E \propto 1/N$ and $\Delta \gamma \propto 1/N$, in Hermitian and PT-symmetric MLLs, respectively. The inversely proportional behaviors are caused by a linear increase of the energy density with system size and the relation between energy $E_{\pm}$ and the parameter $\gamma$ in Eq.~(\ref{eq_band}).

Consequently, if the bulk size is infinite, the boundary effect disappears and thus the CLL and MLL can be associated with the same energy bands irrespective of two different periodic boundary conditions. Comparing this with a circular cavity with a scatterer, the ratio between bulk and boundary size corresponds to the ratio between cavity and scatterer size. The smaller scatterer is compared to the cavity, the smaller is the effect of the scatterer, i.e., the difference between even and odd parity modes.

\subsection{Summary: Möbius ladder lattices and chaotic billiards}
\label{sec_chaoticbilliards}

We can compare our results in the CLL and MLL with level crossings and avoided crossings in regular and chaotic billiards as follows: first, we consider eigenenergies in a regular, for example rectangular, billiards. The eigenenergies will show level crossings as a system parameter (e.g., width of the rectangle) varies because of the symmetry of the rectangular billiards. That is, there are degeneracies. If we add deformation as a perturbation to the geometry, e.g., circular arc of Sinai billiards, the systems are desymmetrized. If the systems are fully desymmetrized, all eigenstates 
exhibit avoided level crossing. As a result, degeneracy is lifted due to perturbation. While level crossings change into avoided crossings around the degeneracy, eigenenergies far from degeneracy almost do not change even though we add the perturbation. 

In our case, first we consider circular ladder lattices and find 
many level crossings and degeneracies because of the symmetry of the system. If we add a perturbation, here the sharp twist of the MLL, we break the symmetry of the system. So if we consider Hermitian asymmetric onsite potentials, we can find the avoided crossings near the degeneracy points such as in the cases of rectangular billiards and Sinai billiards.

Finally, if we consider PT-symmetric onsite potential, we can still break the (geometric) symmetry of the system, but the system can preserve the PT-symmetry. This is thus not a fully desymmetrized system in contrast to the Hermitian case. If we now add the perturbation (sharp twist), the degeneracy is lifted, that is, the broken states occur near the degeneracy points due to the strong perturbation. At the same time, the states far from the degeneracy points will be almost the same as the states in circular ladder lattices, that is, the PT-symmetric states. There is a pair of PT-phase transitions at the left and right of the degeneracy points because there are only two phases, PT-symmetric and broken states, in the system preserving PT-symmetry. 



We end with an outlook on realizing coupled ring systems with nontrivial properties. Using two microdisk cavities, coupled e.g.~evanescently via free space, and featuring loss and gain as well as an asymmetric coupling induced, e.g.~by scatterers, may allow one to reproduce some of the described properties related to the combination of geometric asymmetry and the presence of PT symmetry in a realistic mesoscopic optics system, and may trigger future applications. 

To summarize, we have shown that non-Hermitian properties in mesoscopic systems can arise from asymmetric real scattering or from symmetric complex scattering. Here, it was induced by the topological twist of a Möbius ladder lattice that we found to induce a new type of avoided level crossing in the PT symmetric situation. Using an effective description with 2 $\times$ 2 matrices proved, once more, the usefulness of this approach in gaining a deeper physical understanding of the system. 

\section{Acknowledgements}
We thank Sibylle Gemming for valuable discussions. We acknowledge financial support from the Institute for Basic Science in the Republic of Korea through the project IBS-R024-D1.

\section{References}

\bibliographystyle{unsrt}
\bibliography{bibtex_all.bib}

\end{document}